# Bond Lengths of Single-Walled Carbon Nanotubes

**Ali Nasir Imtani and V. K. Jindal***

**Department of Physics, Panjab University, Changdigrah-160014, India.**

## Abstract

Results of the bond lengths for various chiralities of single-wall carbon carbon nanotubes (SWNTs) (armchair, zigzag and chiral) are obtained. We use modified helical and rotational symmetries to describe the structure of SWNTs and Tersoff potential to minimize the energy of these tubes. It emerges that in general, two bond lengths are required for obtaining minimum energy structure, in contrast to one bond length commonly used. The difference in bond lengths depends on chirality and radius of achiral tubes. Significantly, even a small deviation from zigzag or armchair character leads to interesting behavior of bond lengths. A reduction in diameter is responsible for difference in the bond lengths of achiral nanotubes. We also calculate the bond lengths under hydrostatic pressure. The behavior of bond lengths for armchair single-wall nanotubes when calculated under pressure shows that the larger bond length decreases faster with pressure in comparison to the shorter bond length. At some critical pressure (depending upon the radius of the tube), the two bond lengths become equal to each other, reversing their difference above this critical pressure. We suggest that this behavior can be exploited to experimentally determine the chirality and radius of the carbon nanotubes, for example by observing the presence and disappearance of modes typical of two different bond lengths. This change occurs only within a few GPa of pressure.

## Introduction

Since their discovery in 1991[1], both single-wall and multi-wall nanotubes have become an active area of research. The structure of carbon nanotubes has been explored with the help of transmission electron microscopy (TEM) and scanning tunneling microscopy (STM), yielding direct confirmation that the nanotubes are seamless cylinders derived from the honeycomb lattice of graphite sheet, a single atomic layer of crystalline graphite. We focus on single-wall nanotubes (SWNTs) which were first reported in 1993 [2,3]. The SWNT are characterized by strong covalent bonding, a unique one-dimensional structure and nanometer size which impart unusual properties to the nanotubes including exceptionally high tensile strength, high resilience, electronic properties ranging from metallic to semiconducting, high current carrying capacity, and high thermal conductivity. Many studies have investigated the variations of bond lengths in SWNTs from graphite value. Robertson *et al.* [4] used first principle LDF method to calculate the total energies for a series of high symmetry tubules (n,n). They found that the minimum energy structure of (5,5) tubule by direct minimization of the total energy gives a radius of 3.47Å with carbon-carbon bonds of length of 1.44Å. Several other researchers [5-9] have reported calculations based on using one bond length equivalent to graphite value or a modified value leading to results on many properties of carbon nanotubes like Young's modulus , bulk and structural properties and other lattice and thermodynamical properties. Danil *et al.*[10] have reported the results for armchair single-wall nanotubes based on two bond lengths using *ab intio* calculations. They found that both bond lengths have values greater than the value of bond length of graphite. The investigation of curvature effect on geometric parameters, energy, and electronic structure of zigzag nanotubes using first-principle calculation has also been done by Gulseren *et al.* [11]. They report again that two bond lengths and bond angles display a monotonic variation and approach the graphite values as radius increases. One of the bond lengths which it parallel to tube axis is less than the graphite value whereas the other which make an angle with tube axis is grater than it. Except these [10,11], no calculation seems to use two bond lengths for formulating the structure and properties of carbon nanotubes.

* Author with whom correspondence be made, e-mail: jindal@pu.ac.in

It turn out that whereas in a graphite sheet one bond length determines it structure, a carbon nanotubes involves two bond lengths in general. This is because a graphene on folding incorporates two different directions, a radial direction and a length direction, which are likely to be dictated through two different bond lengths. The extent of difference in bond lengths may vary with chirality and radius. In armchair tubes, there are two possible bond lengths and apparent shown in Fig.1, one bond length $b_1$ perpendicular to tube axis and other $b_2$ making an angle with it. Zigzag tubes also similarly have two bond lengths (see Fig.1), one of them parallel to tube axis. In chiral tubes these both bond lengths make an angle with the tube axis. The different direction of bond lengths with tube axis presents different behavior of bond lengths in the circumference and in axial directions. For this reason one expects that there is different behavior of the bond lengths in three types of the single-wall nanotubes after rolling up the graphite sheet to construct these nanotubes.

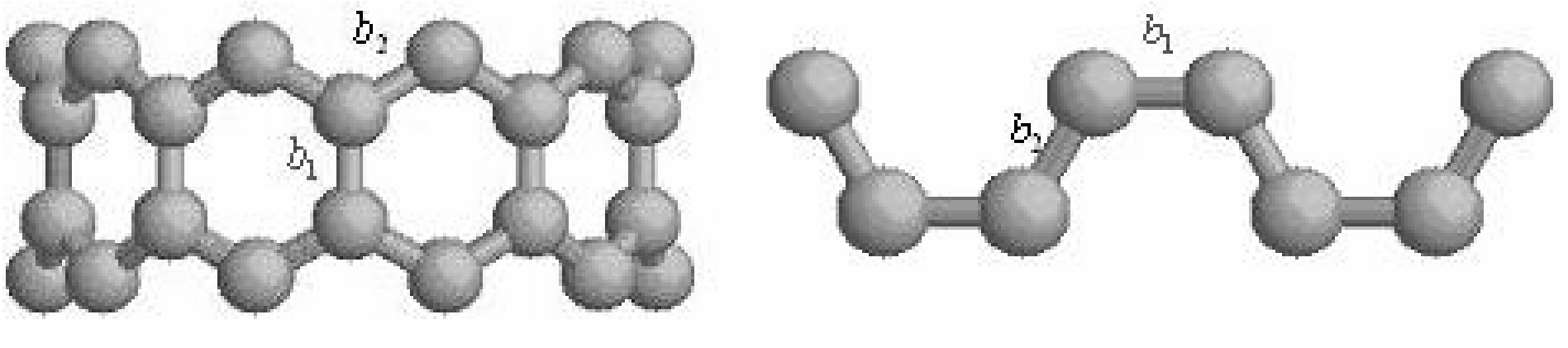

Figure 1 : A part of zigzag (left) and armchair (right) single-wall nanotubes indicating two types of C-C bonds. These are labeled as $b_1$ and $b_2$.

In this paper, we present a comparison of bond lengths for several chirality and several radii carbon nanotubes using modified helical and rotational symmetries to describe the structure of carbon nanotubes [12] and Tersoff potential [13] to minimize the energy of various chirality carbon nanotubes. A subsequent paper will be aimed at exploring these bond lengths features in more detail We compare our results of energy and strain under hydrostatic pressure using thus obtained two bond lengths. We compare our results with those using one bond length also.

## Theoretical Procedure

We can visualize an infinite tube as a conformal mapping of a two-dimensional honeycomb lattice to the surface of a cylinder that is subject to periodic boundaries both around the cylinder and along its axis. The helical and rotational symmetries [12] are used in this study to construct a high symmetry armchair (n,n) and zigzag (n,0) SWNTs and chiral SWNTs as a function of two bond lengths instead of one bond length. As an example, for armchair this is done by first mapping the two atoms in the [0,0] unit cell to the surface of cylindrical shape. The first atom is mapped to an arbitrary point on the cylindrical surface (e.g. $(R,0,0)$ ), where $R$ the tube radius in terms of bond lengths $b_1$ and $b_2$ and the position of the second atom is found by rotating this point by $\phi = 2\pi / 3n$ (radian) about the cylinder axis in conjunction with its translation by $h_s$ (which is for armchair equal to zero) along the tube axis. These first two atoms can be used to locate $2(n-1)$ additional atoms on the cylindrical surface by $(n-1)$ successive $2\pi / n$ rotations about the cylinder axis. Altogether, these 2n atoms complete the specification of the helical motif which corresponds to an area on the cylindrical surface. This helical can then be used to tile the reminder of the tubule by repeated operation of a single screw operation $S(h,\alpha)$ representing a translation $h$ unit along the cylinder axis and rotation $\alpha$ about this axis, where $h = \sqrt{3}b_2 / 2$ and $\alpha = 2\pi / n$ (radian). If we apply the full helical motif, then the entire structure of armchair SWNT is generated. This structure provides atom position of all atoms in terms of bond lengths. The bond lengths are determined by minimization of the energy of the tube, assuming atoms interact via Tersoff potential. For zigzag and chiral tubes there are modified relations of helical screw as a function of bond length $b_1$ or $b_2$ depending on the structure of these tubes [14,15].

## Results and Discussion

By using the minimization energy procedure briefly stated above and detailed elsewhere [14], we obtained a spectrum of these two bond lengths whose difference depend on the chirality and radius. Results of the normalized values of these bond lengths (i.e. $b_{1,2} / b_o$ where $b_o$ is the bond length in graphite sheet) for single-wall nanotubes having different radii are obtained. The results for armchair single wall carbon nanotubes are plotted as a function of radius of tube in Fig. 2. The bond length $b_1$ (see Fig.1) is greater than that of bond length of graphite whereas the bond length $b_2$ is less than that of graphite. The work of Daniel et al [10], who also use two bond lengths, shows different behavior of bond lengths in relation to graphite. As the radius of tube increases the two bond lengths approach each other as well as to the value of bond length of graphite, as expected.

In order to see the influence of two bond lengths in comparison to one bond length, we also use our models for armchair SWNTs to calculate the energy assuming that these are equal (i.e. $b_1 = b_2 = b$). Again, minimization of the energy leads to the values of this bond length. The results for armchair SWNTs are listed in Table 1 and the normalized bond length plotted as a function of the tube radius in Fig. 2 along with the results of tubes having unequal bond lengths. We observe that for these tubes the radius in equal bond length structure is less than that in unequal bond lengths structure for the same tube. As an example, for (5,5) tube the values of two bond lengths $b_1$ and $b_2$ are 1.44 Å, 1.417744 Å and corresponding value of the radius is 3.43774 Å. When the structure with equal bond lengths is considered the value of bond length and corresponding radius comes out to be equal to 1.429 Å and 3.41148 Å, respectively. It is important to note that the minimized energy is lower using two bond lengths as compared to using one bond length. Detailed results can be seen in Table 1. This indicates that two bond length structure is the structure of minimum energy.

We recalculate the bond lengths under hydrostatic pressure on the tube. Again, we take a (5,5) tube for our study. Using one bond length results in equal compression along axial and circumferential directions as shown in Fig. 3a. This contrasts with the results using unequal bond lengths for the same tube, where the compression along the length of tube is less than along the circumference as shown in Figure 3b.

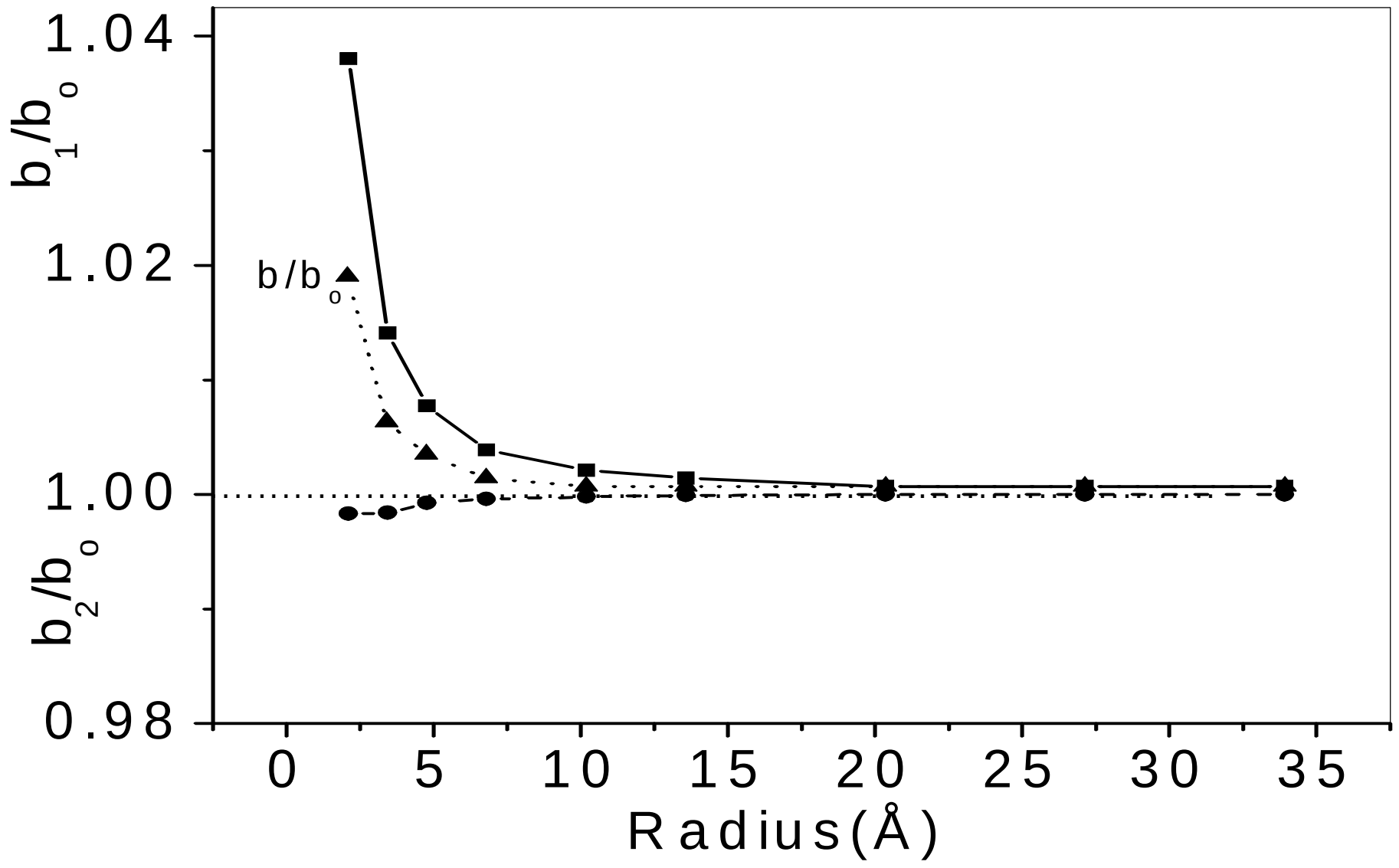

Figure 2: Variations of two bond lengths with the tube radius for armchair (n,n) single-wall nanotubes, where n=3, 5, 7, 10, 15, 20, 30, 40, and 50 having different radii. Solid line shows the behavior of $b_1$ and the broken line that of $b_2$. We also show as dotted line the results using one bond length ($b$) only. The lines are only guides to the eye, results have been calculated for tubes on the points shown.

Table 1: Radius, bond lengths, energy $E(eV/atom)$ of armchair SWNTs assuming the tube structure with equal bond lengths.

| | Equal bond lengths structure | | | Unequal bond lengths structure | | | |
|---|---|---|---|---|---|---|---|
| SWNT | Radius(Å) | $b$ (Å) | $E$ | Radius(Å) | $b_1$ (Å) | $b_2$ (Å) | $E$ |
| (3,3) | 2.07267 | 1.447 | -6.93524 | 2.11134 | 1.474 | 1.41764 | -6.96145 |
| (5,5) | 3.41148 | 1.429 | -7.19112 | 3.43774 | 1.44 | 1.41774 | -7.19580 |
| (7,7) | 4.76271 | 1.425 | -7.25429 | 4.78276 | 1.431 | 1.41894 | -7.25570 |
| (10,10) | 6.78954 | 1.422 | -7.28610 | 6.80626 | 1.425 | 1.41942 | -7.28650 |
| (15,15) | 10.1771 | 1.421 | -7.30253 | 10.1914 | 1.423 | 1.41975 | -7.30263 |
| (20,20) | 13.5695 | 1.421 | -7.30818 | 13.5790 | 1.422 | 1.41991 | -7.30823 |
| (30,30) | 20.3543 | 1.421 | -7.31218 | 20.3543 | 1.421 | 1.41999 | -7.31220 |
| (40,40) | 27.1391 | 1.421 | -7.31358 | 27.1391 | 1.421 | 1.41999 | -7.31359 |
| (50,50) | 33.9238 | 1.421 | -7.31423 | 33.9238 | 1.421 | 1.41999 | -7.31423 |

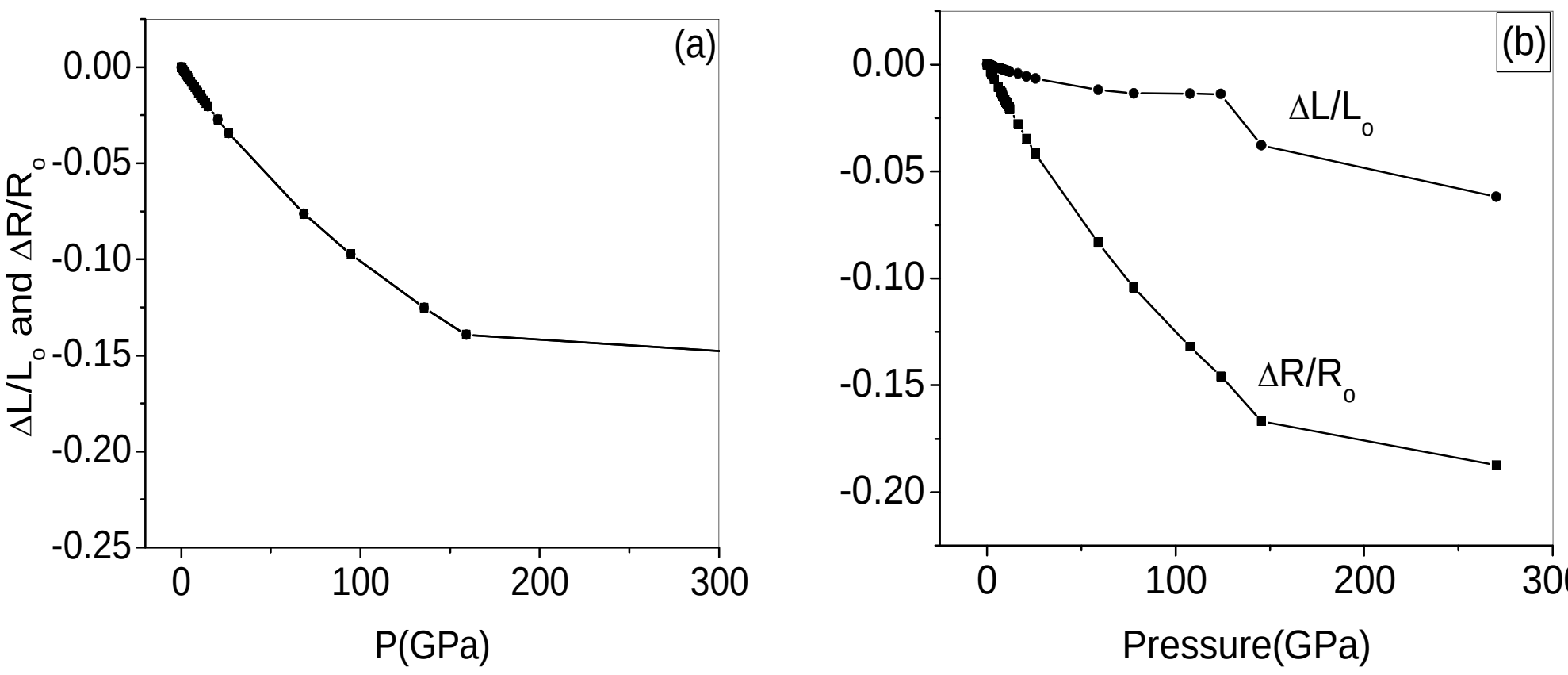

Figure 3: The compression of (5,5) tube along the length and radial directions when a hydrostatic pressure is applied on the tube. Figure (a) shows results from equal bond length tube and (b) for unequal bond length tube.

We now study (5,5), (10,10), (15,15) and (30,30) SWNTs for their bond length behavior under hydrostatic pressure.

The calculated value of the bond lengths under hydrostatic pressure are plotted in Fig. 4. We observe that both bond lengths $b_1$ and $b_2$ decrease under pressure. The larger bond length $b_1$ decreases faster with pressure as compared to the shorter bond length $b_2$. At some critical values of pressure ($P_c$) both bond lengths become equal to each other ($b_1 = b_2 = b_c$). The values of this critical pressure and of critical equal bond lengths ($b_c$) are dependent on the radius of tube. Above this critical pressure, the bond lengths reverse their difference, the shorter bond length changes to bigger bond length and vice versa. We show the variation of $P_c$ with radius as well as $b_c$ in Fig. 5. It also emerges that $P_c$ reduces with increasing radius of the tube. For large radius, $P_c$ approaches to zero and $b_c$ approaches to that for graphite, as expected.

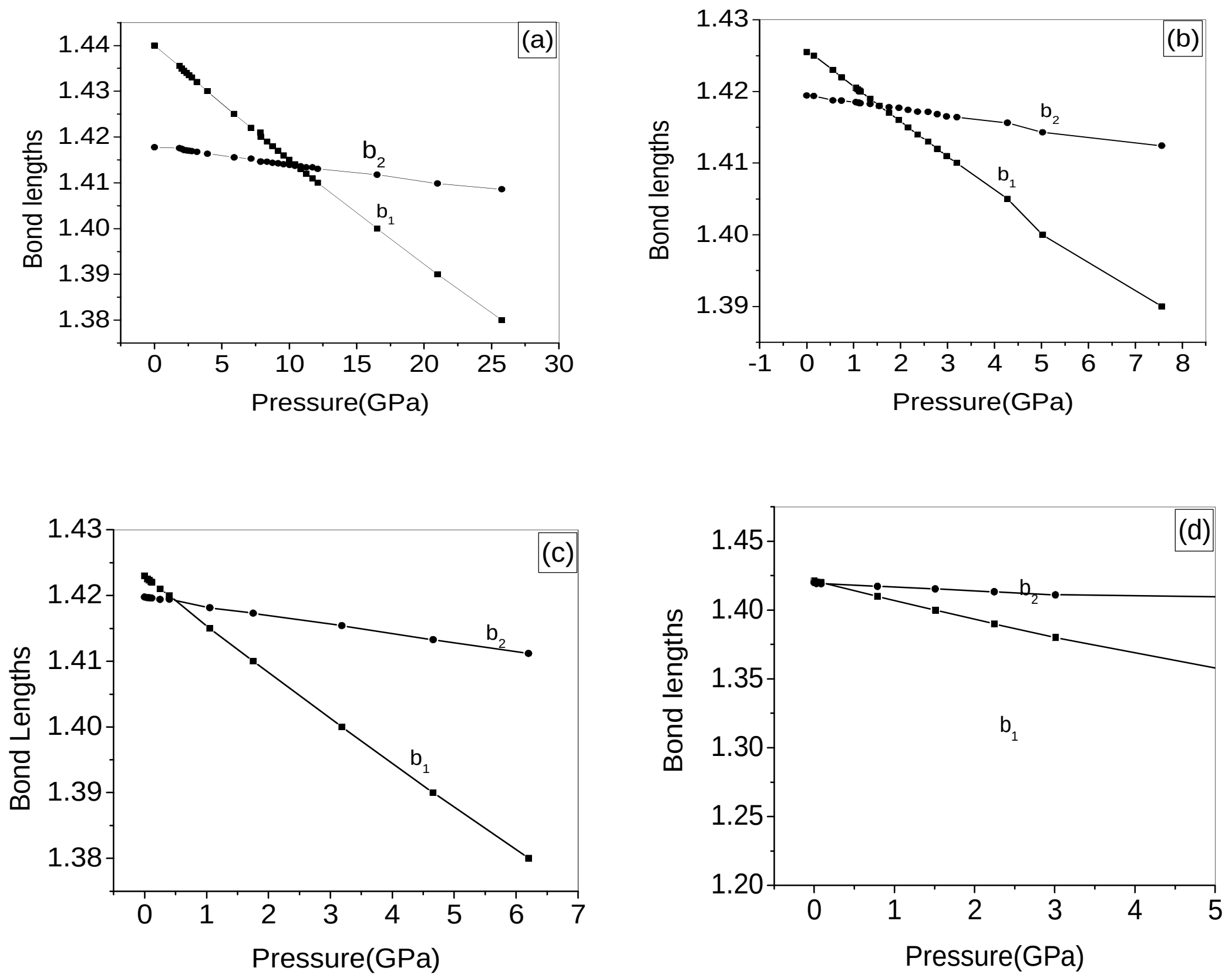

Figure 4: The behavior ob both bond lengths $b_1$ (Å) and $b_2$ (Å) under a hydrostatic pressure for armchair SWNTs. (a) (5,5), (b) (10,10), (15,15) and (30,30) tubes.

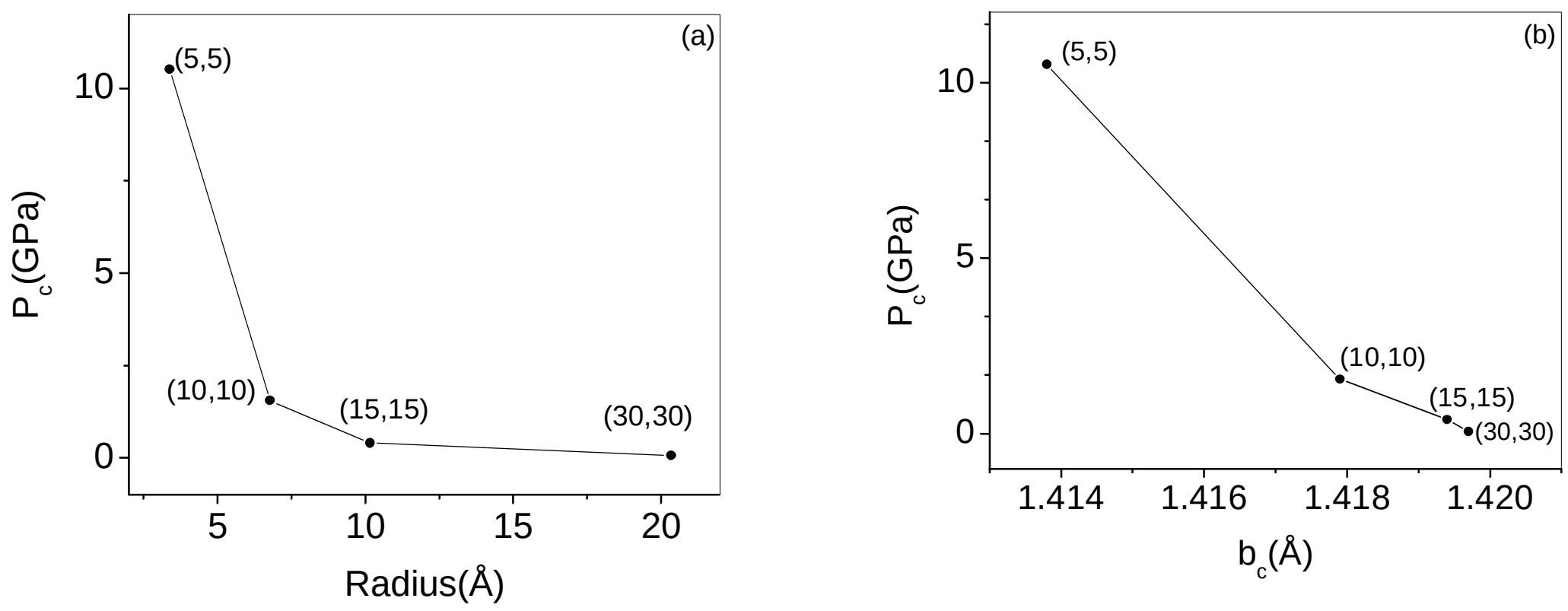

Figure 5 : Critical pressure as a function of (a) radius and (b) bond length $b_c$ .

In zigzag single-wall nanotubes the situation is different from armchair. Here the bond length $b_1$ is parallel to tube axis (Fig.1). The results obtained for these bond lengths are plotted as a function of radius of tube are shown in Fig.6. We observe that the values of bond length $b_1$ are approximately fixed and equal to the value of bond length of the graphite and independent of radius of tube. We also see in the same figure the behavior of bond length $b_2$ which makes an angle with tube axis. The values of this bond length are greater than graphite value and dependent on the tube radius. The details of calculation for zigzag tubes and results under pressure will be published separately [15]. Gulseren *et al.* [11] also seem to use two bond lengths for zigzag tubes, but details of procedure for obtaining their equilibrium values are unclear.

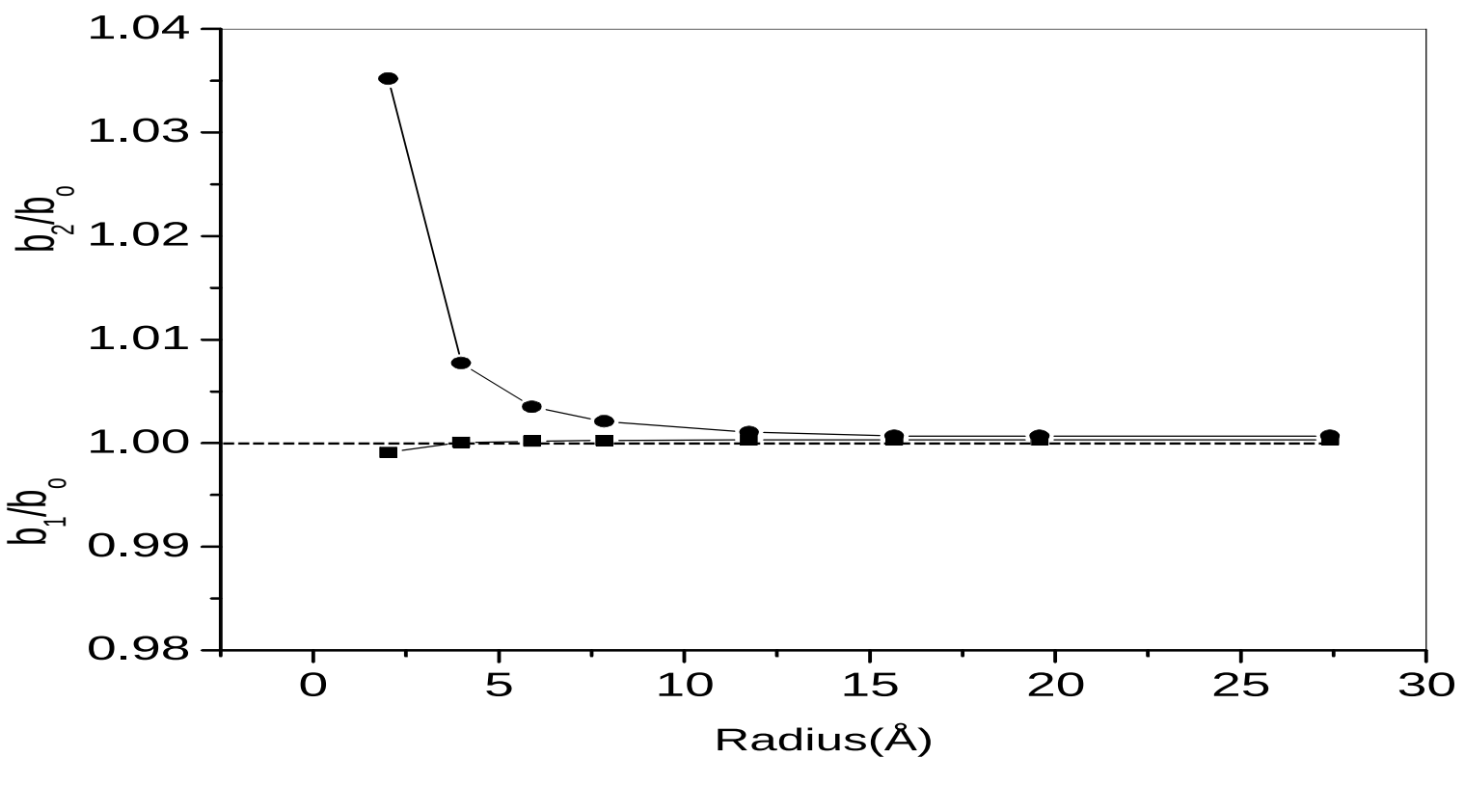

Figure 6: Variations of normalized bond lengths with the tube radius for zigzag (n,0) single-wall nanotubes, where n= 5, 10, 15, 20, 30, 40, 50 and 70 having different radii.

The bond lengths in chiral nanotubes behave differently. Here both bond lengths make an angle with the tube axis. All chiral tubes show minimum energy for $b_1 = b_2 = b$ . Fig.7 shows the results of variations the bond length as a function of radius in different chiral angles. We observe that the values of bond lengths associated with minimum energy are equal for all chiral nanotubes having middle and large radii in the same chiral angle. These values of bond length depend on the radius only for small radius tube.

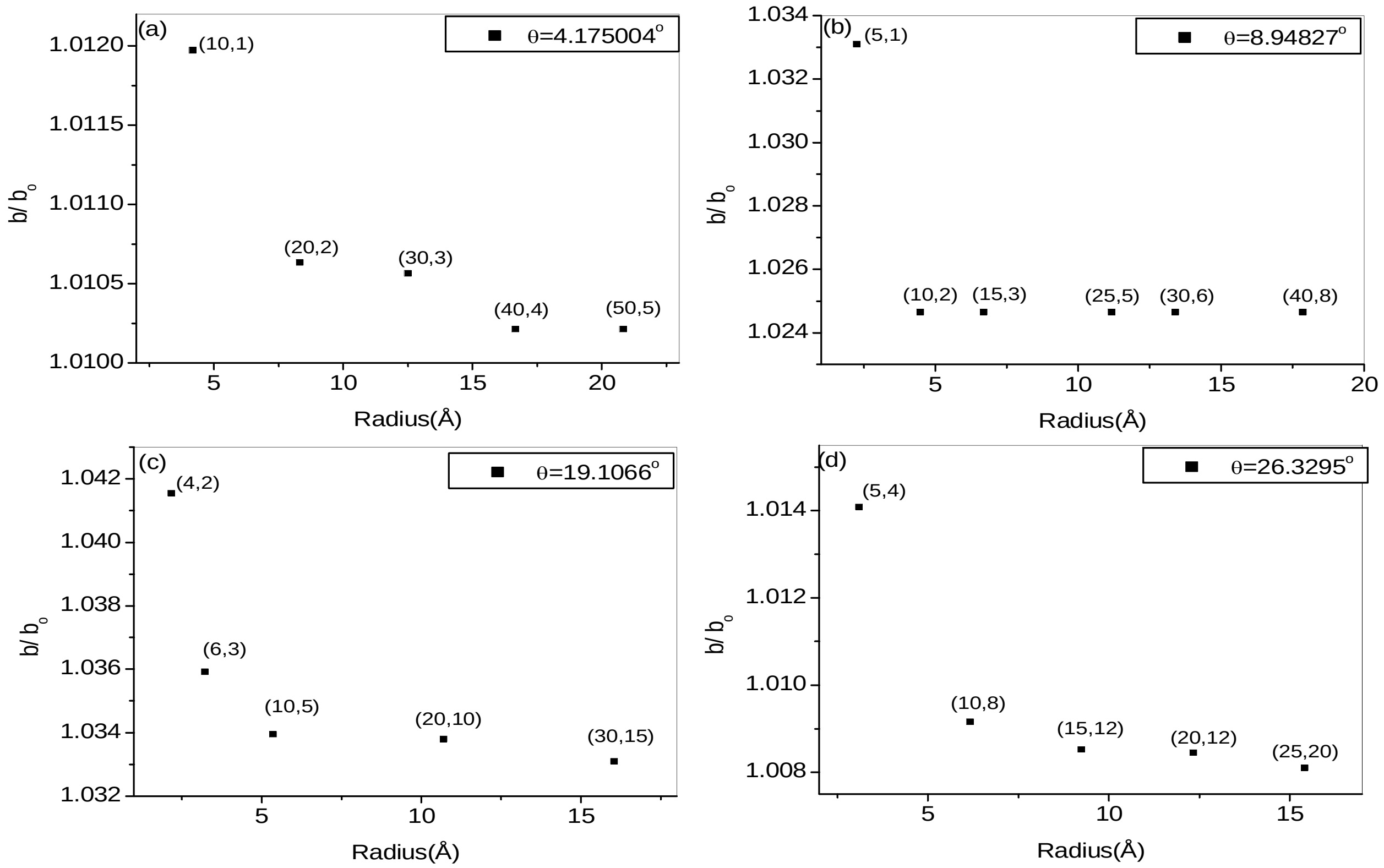

Figure 7: Variations of normalized bond length with tube radius at various chiral angles.

# Summary and Conclusion

In this paper we investigate the variations of bond lengths with tube radius and chiral angle and compare results for all types of single-wall nanotubes using a procedure as described briefly in this study and in detail in our earlier work [14]. It turns out that generally two bond lengths are required to properly obtain a minimized structure, in contrast with one bond length used extensively in the literature uptil now. The two bonds have different roles in obtaining the overall radius and length of a given tube. For this reason we expect the single-wall nanotubes to have different properties especially those which differentiate length from diameter. As a result of this work, formation of tube from a graphene results in differences in bond lengths, these differences being characteristic of the diameter size and chirality of the tubes so formed. A smaller diameter tube presents more pronounced difference. We also subject these tubes to hydrostatic pressure, and resulting analysis is much more interesting and significant. The two bond lengths compress differently and interchange from shorter to longer bond length and vice versa at some easily achievable critical pressure values. We feel that these effects in bond lengths can be observed by experimentally in the form of new intra-tube modes in comparison to the ones using one bond length. An appearance or disappearance of a mode under hydrostatic pressure from a given chirality and radius tube can be exploited to provide useful structural information. It seems work similar to Merlen et al.[16] can be very useful for this purpose. As a result, we observe a pattern in bond length variation with curvature and chirality.

The present analysis underlines why strong curvature obtained by nano-sized diameter have different bulk, structure and thermodynamical properties as compared to a planner graphite sheet. It also proposes new techniques to determine experimentally the chirality and diameter of SWNT.

# References

[1] S. Iijima, Nature, 354, 56 (1991).

[2] D. S. Betune, C. H. Kiang, M. S. de Vries, G. Gorman, R. Savoy, J. Vazquez, R. Bevers, 'Cobalt-Canalized Growth of Carbon Nanotubes with Single-layer Walls', Nature, 363,605 (1993).

[3] F. Kokai, K. Takahashi, M. Yudasaka, R. Yamada, T. Ichihashi, S. Iijima, 'Growth Dynamics of Single-wall Carbon Nanotubes Synthesized by $CO_2$ Laser Vaporization', J. Phys. Chem. B 103, 4345 (1999).

[4] D. H. Robertson, D. W. Brenner and J. W. Mintmire, ' Energetics of nanoscale graphitic tubules', Phys. Rev. B 45, 12592(1992-I).

[5] Guanghua Gao, Tahir Cagin and William A Goddard III, Nanotechnology 9,184 (1998).

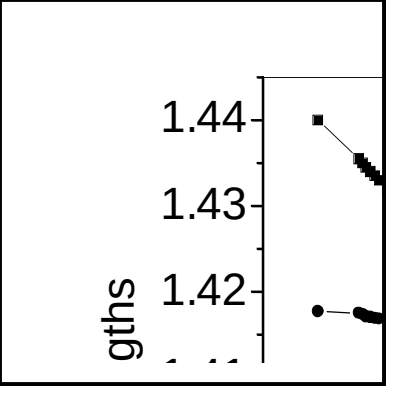

[6] Maria Huhtala, Antti Kuronen and Kimmo Kaski, Comput. Phys. Commum. 147, 91 (2002).

[7] Shanker Ghosh, Pallavi V. Teredesai and A. K. Sood,Pure Appl. Chem. 74, 1719 (2002).

[8] A. Sears and R. C. Batra, Phys. Rev. B 69, 235406 (2004).

[9] Danil Sanclez-Portat, Emilio Artacho, Jose M. Soler, Angel Rubio and Pablo Ordejon, Phys. Rev. B 59, 12678(1999).

[9] Min Ouyang, Jin-Lin Huang and Charles M. Lieber, Acc. Chem. Res. 35, 1018 (2002).

[11] O.Gulseren, T. Yildirim and S. Ciraci, 'Systematic ab initio study of curvature effects in carbon nanotubes', J. Phys. Rev. B 65, 153405(2002).

[12] C. T. White, D. H. Robertson and J. W. Mintmire,'Helical and rotational symmetries of nanoscale graphitic tubules', J. Phys. Rev. B 47, 5485(1993).

[13] J. Tersoff, 'New empirical approach for the structure and energy of covalent systems', Phys. Rev. B 37, 6991(1988).

[14] Ali Nasir Imtani and V.K. Jindal submitted to Phys. Rev. B, (2006).

[15] Ali Nasir Imtani and V. K. Jindal, to be published.

[16] A. Merlen, N. Bendiab, P. Toulemonde, A. Aouizerat, A. San Miquel, J. L. Sauvajol, G. Montagnac, H. Cardon and P. Petit, Phys. Rev. B 72, 035409 (2005).